\documentclass[format=acmsmall, review=false, screen=true]{acmart}

\usepackage{booktabs} 
\usepackage{amsmath}
\usepackage{graphicx}
\usepackage{epstopdf}
\usepackage{epsfig}
\usepackage{graphics}
\usepackage{array}
\usepackage{float}
\usepackage{multirow}

\usepackage[ruled]{algorithm2e} 

\SetAlFnt{\small}
\SetAlCapFnt{\small}
\SetAlCapNameFnt{\small}
\SetAlCapHSkip{0pt}
\IncMargin{-\parindent}

\acmJournal{CSUR}
\acmVolume{-}
\acmNumber{-}
\acmArticle{-}
\acmYear{2023}
\acmMonth{04}

\providecommand{\tabularnewline}{\\}
\floatstyle{ruled}
\newfloat{algorithm}{tbp}{loa}
\providecommand{\algorithmname}{Algorithm}
\floatname{algorithm}{\protect\algorithmname}

\makeatother

\begin{document}
\bibliographystyle{acmtrans}
\title{A Comprehensive Survey on the Implementations, Attacks, and Countermeasures of the
Current NIST Lightweight Cryptography Standard}

\author{Jasmin Kaur}
\affiliation{%
  \institution{University of South Florida}
  \city{Tampa}
  \state{FL}
  \postcode{33620}
  \country{USA}}
\email{jasmink1@usf.edu}
\author{Alvaro Cintas Canto}
\affiliation{%
  \institution{Marymount University}
  \city{Arlington}
  \state{VA}
  \postcode{22207}
  \country{USA}}
\email{acintas@marymount.edu}
\author{Mehran Mozaffari Kermani}
\affiliation{%
  \institution{University of South Florida}
  \city{Tampa}
  \state{FL}
  \postcode{33620}
  \country{USA}}
\email{mehran2@usf.edu}
\author{Reza Azarderakhsh}
\affiliation{%
  \institution{Florida Atlantic University}
  \city{Boca Raton}
  \state{FL}
  \postcode{33431}
  \country{USA}}
\email{razarderakhsh@fau.edu}

\begin{abstract}
This survey is the first work on the current standard for lightweight
cryptography, standardized in 2023. Lightweight cryptography plays
a vital role \textcolor{black}{in }securing resource-constrained embedded
systems such as deeply-embedded systems (implantable and wearable
medical devices, smart fabrics, smart homes, and the like), radio
frequency identification (RFID) tags, sensor networks, and privacy-constrained
usage models. National Institute of Standards and Technology (NIST)
initiated a standardization process for lightweight cryptography and
after a relatively-long multi-year effort, eventually, in Feb. 2023,
the competition ended with ASCON as the winner. This lightweight cryptographic
standard will be used in deeply-embedded architectures to provide\textcolor{red}{{}
}\textcolor{black}{security through confidentiality and integrity/authentication}\textcolor{red}{{}
}\textcolor{black}{(the dual of the legacy AES-GCM block cipher which
is the NIST standard for symmetric key cryptography).} ASCON's lightweight
design utilizes a $320$-bit permutation which is bit-sliced into
five $64$-bit register words, providing $128$-bit level security.
This work summarizes the different implementations of ASCON on field-programmable
gate array (FPGA) and ASIC hardware platforms on the basis of area,
power, throughput, energy, and efficiency overheads. The presented
work also reviews various differential and side-channel analysis attacks
(SCAs) performed across variants of ASCON cipher suite in terms of
algebraic, cube/cube-like, forgery, fault injection, and power analysis
attacks as well as the countermeasures for these attacks. We also
provide our insights and visions throughout this survey to provide
new future directions in different domains. This survey is the first
one in its kind and a step forward towards scrutinizing the advantages
and future directions of the NIST lightweight cryptography standard
introduced in 2023.
\end{abstract}


 \maketitle

\section{Introduction}

Lightweight cryptography (LWC) has become a necessity today as the
world is extensively adopting the Internet of Things (IoT), and the
Internet of Nano-Things. LWC is extensively used in resource constraint
devices such as radio frequency identification (RFID) tags, wireless
sensor networks (WSN), and embedded systems (implantable and wearable
medical devices, smart fabrics, smart homes, and the like) to ensure
their applications are secure. However, security does not mean reliability,
and many lightweight cryptographic algorithms such as ASCON are vulnerable
to SCAs. Before any standardization efforts existed for lightweight
cryptography, there were many research work performed on various aspects
of lightweight ciphers including the works by the authors, e.g., for
side-channel analysis (SCA) \cite{key-41}-\cite{key-47}.

The National Institute of Standards and Technology (NIST) initiated
a standardization process for lightweight cryptography and after a
relatively-long multi-year effort, eventually, in Feb. 2023 the competition
ended with ASCON \cite{key-1} as the winner among the other round
three candidates - Elephant \cite{key-48}, GIFT-COFB \cite{key-49},
Grain128-AEAD \cite{key-50}, ISAP \cite{key-51}, Photon-Beetle \cite{key-52},
Romulus \cite{key-53}, Sparkle \cite{key-54}, TinyJambu \cite{key-55},
and Xoodyak \cite{key-56}. This lightweight cryptographic standard
will be used in deeply-embedded architectures to provide security.
Previously, ASCON was also chosen as a finalist of the CAESAR competition
for authenticated encryption. 

ASCON is a lightweight cipher suite that provides authenticated encryption
with associated data (AEAD) as well as hashing functionalities. It
uses a duplex-based mode of operation \cite{key-1}. The $320$-bit
permutation of ASCON iteratively applies a substitution-permutation
network to encrypt/decrypt data in a bit-slice fashion. This bit-slice
implementation of ASCON permutation makes it scalable to $8$-, $16$-,
$32$-, and $64$-bit platforms while remaining lightweight. ASCON
has two different variants for different message lengths - ASCON-128
and ASCON-128a. ASCON-128 uses a message length of $64$ bits while
ASCON-128a uses a message length of $128$ bits. ASCON also has a
post-quantum secure variant called ASCON-128pq which is the same as
ASCON-128 but uses a $160$-bit length key \cite{key-1}. We note
that post-quantum cryptography (PQC) refers to attacks enabled at
the presence of powerful quantum computers. The algorithms for public-key
cryptography were standardized in 2022; yet, symmetric-key cryptography
has much less issues and the larger keys can be used for such threats.
The current NIST PQC winners are CRYSTALS-KYBER \cite{key-57}, CRYSTALS-DILITHIUM
\cite{key-58}, FALCON \cite{key-59}, and SPHINCS+ \cite{key-60}.

Over the years, various cryptanalysis, both differential and side-channel,
have been performed on different ASCON variants. Madushan et al. \cite{key-2}
explore the various fault analysis of the NIST LWC standardization
process finalists. Furthermore, Dobraunig et al. \cite{key-3} perform
in- depth cryptanalysis of ASCON for key-recovery attacks, forgery
attacks, and algebraic attacks by using zero-sum distinguisher. Moreover,
they leverage the low algebraic degree of ASCON to construct a zero-sum
distinguisher, i.e., a set of input and output values for which sum
to zero over $\mathbb{\mathbb{F}}_{2}^{n}$, for the $12$-round ASCON
that is able to highlight the ASCON permutation from a random permutation
with a complexity of $2^{130}$ by targeting the internal state after
round $5$. The recent cryptanalysis of ASCON has strived towards
improving the work of \cite{key-3} as well as to propose new methodology
for determining new distinguishers for differential, cube, algebraic,
and forgery based key-recovery attacks. This study extends the work
of \cite{key-2}, and summarizes the new differential cryptanalysis
and SCA works performed on ASCON in hardware/software implementations. 

The SCA works reviewed in this survey also include statistical fault
analysis (SFA), machine learning (ML), and differential power analysis-based
key-recovery attacks performed on the hardware implementations of
ASCON. In SFA, the attacker performs a statistical analysis of the
injected fault on the output of the ASCON permutation to fully recover
the secret key;\textcolor{red}{{} }\textcolor{black}{in machine learning
(ML) strategies implement deep and reinforced/unsupervised learning
techniques to extract the secret key; while, the differential power
analysis (DPA) exploit the vulnerabilities in the ASCON initialization
operation to mount key-recovery attacks on the hardware implementations
of ASCON.}\textcolor{red}{{} }The SCA countermeasures, which are also
reviewed in this work, include threshold implementation strategies,
stronger S-box design, error-detection mechanisms, as well as protected
architectures against side-channel leakage. This paper also summarizes
the various hardware implementations of ASCON that have improved the
design for better area utilization, power consumption, throughput,
energy, and efficiency on FPGA and ASIC hardware platforms. We also
provide our insights and visions throughout this survey to provide
new future directions in different domains. This survey is the first
one in its kind and a step forward towards scrutinizing the advantages
and future directions of the NIST lightweight cryptography standard
introduced in 2023.

The organization of the paper is as follows. Section 2 describes the
design and architecture of ASCON. Section 3 explores the hardware
implementations of ASCON presented in previous and current literature.
Section 4 lists the existent differential cryptanalysis and SCAs performed
on ASCON. Finally, we conclude the review in Section 5.

\begin{figure*}[t]
\begin{centering}
\includegraphics[scale=0.9]{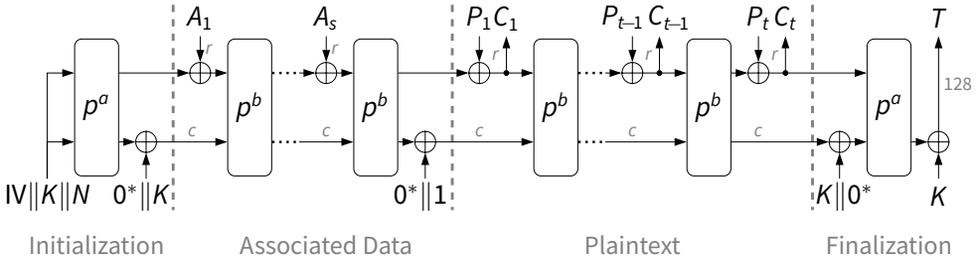}
\par\end{centering}
\caption{The associated encryption of ASCON \cite{key-1}.}
\end{figure*}

\begin{figure}[b]
\begin{centering}
\includegraphics[scale=0.35]{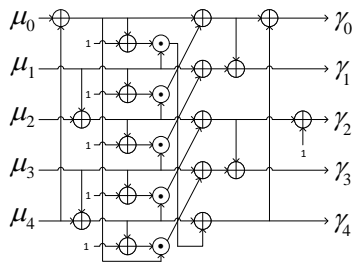}
\par\end{centering}
\caption{The 5-bit S-box of ASCON \cite{key-1,key-40}. }
\end{figure}

\section{Preliminaries}

The entire design specification of ASCON is given in \cite{key-1}.
ASCON's encryption process (Fig. 1) is designed as a sponge-based
MonkeyDuplex construction which consists of $4$ stages, namely, initialization
operation, processing of the associated data, processing of the plaintext,
and the finalization operation. These four stages get updated using
two $320$-bit permutations, i.e., $p^{a}$ and $p^{b}$, where $a$
and $b$ are the number of rounds. Both of the permutations are bit-sliced
into five $64$-bit register words which make up the $5$-bit internal
state. In a full $12$-round ASCON, the permutations iteratively apply
a substitution-permutation network (SPN)-based round transformation
which consists of adding round constants, applying the substitution
layer, and employing the linear layer for diffusion to the internal
state.

The substitution layer consists of a non-linear $5$-bit S-box (Fig.
2) whose hexadecimal form is shown in Table 1. This S-box is applied
$64$ times in parallel to update each bit-slice of the internal state.
The $5$-bit S-box is designed using Boolean logic which makes it
highly compact and lightweight for implementations on both ASIC and
FPGA hardware platforms. The linear layer of ASCON updates each $64$-bit
word of the internal state by first rotating register words with different
shift values, and then performing a modulo-$2$ addition on the shifted
word values.

\begin{table}[t]
\begin{centering}
{\scriptsize{}\caption{LUT representation of the non-linear 5-bit S-box $SB$ of ASCON-128
in hexadecimal form for input vector $\mu$}
}{\scriptsize\par}
\par\end{centering}
\centering{}{\scriptsize{}\setlength\tabcolsep{3.75pt}{}%
\begin{tabular}{|c|>{\centering}m{0.2cm}|>{\centering}m{0.2cm}|>{\centering}m{0.2cm}|>{\centering}m{0.2cm}|>{\centering}m{0.2cm}|>{\centering}m{0.2cm}|>{\centering}m{0.2cm}|>{\centering}m{0.2cm}|>{\centering}m{0.2cm}|>{\centering}m{0.2cm}|>{\centering}m{0.2cm}|>{\centering}m{0.2cm}|>{\centering}m{0.2cm}|>{\centering}m{0.2cm}|>{\centering}m{0.2cm}|>{\centering}m{0.2cm}|}
\hline 
{\footnotesize{}$\mu$} & {\footnotesize{}0} & {\footnotesize{}1} & {\footnotesize{}2} & {\footnotesize{}3} & {\footnotesize{}4} & {\footnotesize{}5} & {\footnotesize{}6} & {\footnotesize{}7} & {\footnotesize{}8} & {\footnotesize{}9} & {\footnotesize{}a} & {\footnotesize{}b} & {\footnotesize{}c} & {\footnotesize{}d} & {\footnotesize{}e} & {\footnotesize{}f}\tabularnewline
\hline 
{\footnotesize{}$SB(\mu)$} & {\footnotesize{}4} & {\footnotesize{}b} & {\footnotesize{}1f} & {\footnotesize{}14} & {\footnotesize{}1a} & {\footnotesize{}15} & {\footnotesize{}9} & {\footnotesize{}2} & {\footnotesize{}1b} & {\footnotesize{}5} & {\footnotesize{}8} & {\footnotesize{}12} & {\footnotesize{}1d} & {\footnotesize{}3} & {\footnotesize{}6} & {\footnotesize{}1c}\tabularnewline
\hline 
\hline 
{\footnotesize{}$\mu$} & {\footnotesize{}10} & {\footnotesize{}11} & {\footnotesize{}12} & {\footnotesize{}13} & {\footnotesize{}14} & {\footnotesize{}15} & {\footnotesize{}16} & {\footnotesize{}17} & {\footnotesize{}18} & {\footnotesize{}19} & {\footnotesize{}1a} & {\footnotesize{}1b} & {\footnotesize{}1c} & {\footnotesize{}1d} & {\footnotesize{}1e} & {\footnotesize{}1f}\tabularnewline
\hline 
{\footnotesize{}$SB(\mu)$} & {\footnotesize{}1e} & {\footnotesize{}13} & {\footnotesize{}7} & {\footnotesize{}e} & {\footnotesize{}0} & {\footnotesize{}d} & {\footnotesize{}11} & {\footnotesize{}18} & {\footnotesize{}10} & {\footnotesize{}c} & {\footnotesize{}1} & {\footnotesize{}19} & {\footnotesize{}16} & {\footnotesize{}a} & {\footnotesize{}f} & {\footnotesize{}17}\tabularnewline
\hline 
\end{tabular}{\scriptsize{}}}{\scriptsize\par}
\end{table}

\section{Hardware Implementation of ASCON}

This section goes over the various hardware implementations
of the ASCON family along with any optimizations that have been proposed
in recent years. All the overhead results in terms of area, power,
delay, throughput for FPGA implementations, and energy utilization
for ASIC implementations are tabulated in Table 2 and Table 3, respectively.

Various hardware designs of ASCON are implemented in \cite{key-4}
for applications such as RFID tags, WSNs, and embedded systems. Such
hardware implementations of ASCON that \cite{key-4} proposes are:
ASCON-fast, ASCON-64-bit, and ASCON-x-low-area. 

ASCON-fast \cite{key-4} is a high throughput design with minimal
processing delay, which uses unrolled round transformations. At least
one round transformation is performed each clock cycle without any
pipelining. This allows multiple rounds to complete in a single clock
cycle and each ASCON-fast variant uses a different number of the unrolled
round transformations. The unrolled round transformation is connected
with the data bus and key registers using a few additional multiplexers
and XOR gates. 

ASCON-64-bit \cite{key-4} uses an arithmetic logic unit (ALU) where
the control path executes similarly to a sequential code. The design
uses two temporary registers in addition to the five state registers
which along the inputs from the control path constitute the inputs
to the ALU. The ALU takes the 64-bit data input and arranges them
in either the high or low part of the selected operand using a barrel-shift
unit, a data storage unit, and a three logic operations. The result
of the operation is selected at the output of the ALU which is then
applied to the destination register. The S-box and the linear layer
are iteratively calculated using the ALU operations during the execution
phase, thus making one round operation 59 clock cycles. The design
of the S-box is altered to use twenty five 3-operand instructions
and two temporary registers to decrease the area.

In the ASCON-x-low-area variant proposed in \cite{key-4}, the datapath
is designed to use a radical low-area ``one-bit operation per cycle''
approach. The five state registers are clock-gated shift registers
with independent shift-enable inputs. All the state registers are
active during the S-box calculation and the data is shifted bit-slice-wise
in each S-box instance in $64$ clock cycles. The linear diffusion
layer updates each state register individually in five interleaved
sub-iterations. A temporary shift register is used to store the results
of the current linear layer in one iteration and which are then written
back in the next iteration. This low-area design uses $512$ clock
cycles per round transformation. All the overhead results for the
aforementioned implementations are tabulated in Table 3.

Diehl et. al. \cite{key-5} compare the protected and unprotected
implementation of ASCON against first-order differential power analysis
(DPA) using test vector leakage assessment (TVLA) implemented using
Flexible Opensource workBench fOr Side-channel analysis (FOBOS). The
overhead results of the protected ASCON implementations are presented
in Table 2. 

FOBOS 2, an upgraded and optimized FOBOS, is proposed in \cite{key-6}
which is used to evaluate power measurements and SCA resistance for
the hardware implementations of various lightweight ciphers with AEAD
functionality on the Xilinx Artix-7 FPGA board. The results of power
consumption, frequency, throughput, and energy/bit obtained using
FOBOS 2 are tabulated in Table 2. The results show that ASCON performed
better in terms of having the lowest power consumption (33.5 mW at
50 MHz), and lowest incrementally increasing dynamic power with increasing
frequency among the NIST standardization process candidates Spoc,
Spook, and GIFT-COFB. The energy per bit of ASCON was 0.86 nJ/bit
while the static power consumption was around 27 mW.

\begin{table*}[t]
\caption{Overhead results of different hardware implementations of ASCON on
FPGA hardware platforms}

\centering{}{\scriptsize{}\setlength\tabcolsep{4.75pt}{}%
\begin{tabular}{|>{\centering}p{4cm}|>{\centering}p{1.8cm}|>{\centering}p{1.5cm}|>{\centering}p{1.2cm}|>{\raggedright}p{1.5cm}|>{\raggedright}p{1.8cm}|}
\hline 
{\scriptsize{}ASCON }{\scriptsize\par}

{\scriptsize{}Architecture} & {\scriptsize{}FPGA Hardware Platform} & {\scriptsize{}Area }{\scriptsize\par}

{\scriptsize{}(LUTs)} & \begin{centering}
{\scriptsize{}Power }{\scriptsize\par}
\par\end{centering}
\centering{}{\scriptsize{}($mW$)} & \centering{}{\scriptsize{}Throughput (Mbps)} & \centering{}{\scriptsize{}Efficiency (Mbps/LUT)}\tabularnewline
\hline 
\hline 
{\scriptsize{}ASCON (original) \cite{key-40}} & {\scriptsize{}Spartan 7} & {\scriptsize{}371 (slices)} & \centering{}{\scriptsize{}99 } & \centering{}{\scriptsize{}6.646} & \centering{}{\scriptsize{}17.914}\tabularnewline
\hline 
{\scriptsize{}ASCON (original) \cite{key-40}} & {\scriptsize{}Kintex 7} & {\scriptsize{}376 (slices)} & \centering{}{\scriptsize{}88 } & \centering{}{\scriptsize{}6.709} & \centering{}{\scriptsize{}17.843}\tabularnewline
\hline 
{\scriptsize{}ASCON (unprotected) \cite{key-39}} & {\scriptsize{}Spartan 6} & {\scriptsize{}2048} & \centering{}{\scriptsize{}11.5 } & \centering{}{\scriptsize{}255.4} & \centering{}{\scriptsize{}0.1247}\tabularnewline
\hline 
{\scriptsize{}ASCON \cite{key-6}} & {\scriptsize{}Artix7} & {\scriptsize{}1808} & \centering{}{\scriptsize{}26.8} & \centering{}{\scriptsize{}39.0} & \centering{}{\scriptsize{}-}\tabularnewline
\hline 
{\scriptsize{}ASCON-128 \cite{key-7}} & {\scriptsize{}Artix7} & {\scriptsize{}1330} & \centering{}{\scriptsize{}31} & \centering{}{\scriptsize{}457} & \centering{}{\scriptsize{}0.343 }\tabularnewline
\hline 
\multirow{1}{4.3cm}{\centering{}{\scriptsize{}RECO-HCON (128) \cite{key-17}}} & {\scriptsize{}Artix7} & {\scriptsize{}1548} & \centering{}{\scriptsize{}-} & \centering{}{\scriptsize{}$5926$} & \centering{}{\scriptsize{}-}\tabularnewline
\hline 
{\scriptsize{}RECO-HCON (128a) \cite{key-17}} & {\scriptsize{}Artix7} & {\scriptsize{}1548} & \centering{}{\scriptsize{}-} & \centering{}{\scriptsize{}$9077$} & \centering{}{\scriptsize{}-}\tabularnewline
\hline 
{\scriptsize{}RECO-HCON (hash) \cite{key-17}} & {\scriptsize{}Artix7} & {\scriptsize{}1548} & \centering{}{\scriptsize{}-} & \centering{}{\scriptsize{}$3160$} & \centering{}{\scriptsize{}-}\tabularnewline
\hline 
{\scriptsize{}RECO-HCON (hash-a) \cite{key-17}} & {\scriptsize{}Artix7} & {\scriptsize{}1548} & \centering{}{\scriptsize{}-} & \centering{}{\scriptsize{}$4534$} & \centering{}{\scriptsize{}-}\tabularnewline
\hline 
{\scriptsize{}ASCON (protected) \cite{key-39}} & {\scriptsize{}Spartan 6} & {\scriptsize{}6364} & \centering{}{\scriptsize{}37.5} & \centering{}{\scriptsize{}134.6} & \centering{}{\scriptsize{}0.0212}\tabularnewline
\hline 
{\scriptsize{}ASCON-128 (Logic One-bit) \cite{key-40}} & {\scriptsize{}Spartan 7} & {\scriptsize{}373 (slices)} & \centering{}{\scriptsize{}99} & \centering{}{\scriptsize{}6705} & \centering{}{\scriptsize{}17.783}\tabularnewline
\hline 
{\scriptsize{}ASCON-128 (Logic Interleaved-bit) \cite{key-40}} & {\scriptsize{}Spartan 7} & {\scriptsize{}380 (slices)} & \centering{}{\scriptsize{}99} & \centering{}{\scriptsize{}6687} & \centering{}{\scriptsize{}17.444}\tabularnewline
\hline 
{\scriptsize{}ASCON-128 (Logic CRC-3) \cite{key-40}} & {\scriptsize{}Spartan 7} & {\scriptsize{}407 (slices)} & \centering{}{\scriptsize{}99} & \centering{}{\scriptsize{}6603} & \centering{}{\scriptsize{}15.601}\tabularnewline
\hline 
{\scriptsize{}ASCON-128 (LUT One-bit) \cite{key-40}} & {\scriptsize{}Spartan 7} & {\scriptsize{}372 (slices)} & \centering{}{\scriptsize{}100} & \centering{}{\scriptsize{}6448} & \centering{}{\scriptsize{}17.333}\tabularnewline
\hline 
{\scriptsize{}ASCON-128 (LUT Interleaved-bit) \cite{key-40}} & {\scriptsize{}Spartan 7} & {\scriptsize{}377 (slices)} & \centering{}{\scriptsize{}100} & \centering{}{\scriptsize{}6443} & \centering{}{\scriptsize{}17.090}\tabularnewline
\hline 
{\scriptsize{}ASCON-128 (LUT CRC-3) \cite{key-40}} & {\scriptsize{}Spartan 7} & {\scriptsize{}425 (slices)} & \centering{}{\scriptsize{}100} & \centering{}{\scriptsize{}6431} & \centering{}{\scriptsize{}15.131}\tabularnewline
\hline 
{\scriptsize{}ASCON-128 (Logic One-bit) \cite{key-40}} & {\scriptsize{}Kintex 7} & {\scriptsize{}381 (slices)} & \centering{}{\scriptsize{}89} & \centering{}{\scriptsize{}6705} & \centering{}{\scriptsize{}17.598}\tabularnewline
\hline 
{\scriptsize{}ASCON-128 (Logic Interleaved-bit) \cite{key-40}} & {\scriptsize{}Kintex 7} & {\scriptsize{}384 (slices)} & \centering{}{\scriptsize{}89} & \centering{}{\scriptsize{}6687} & \centering{}{\scriptsize{}17.414}\tabularnewline
\hline 
{\scriptsize{}ASCON-128 (Logic CRC-3) \cite{key-40}} & {\scriptsize{}Kintex 7} & {\scriptsize{}385 (slices)} & \centering{}{\scriptsize{}89} & \centering{}{\scriptsize{}6603} & \centering{}{\scriptsize{}17.150}\tabularnewline
\hline 
{\scriptsize{}ASCON-128 (LUT One-bit) \cite{key-40}} & {\scriptsize{}Kintex 7} & {\scriptsize{}363 (slices)} & \centering{}{\scriptsize{}89} & \centering{}{\scriptsize{}6776} & \centering{}{\scriptsize{}18.667}\tabularnewline
\hline 
{\scriptsize{}ASCON-128 (LUT Interleaved-bit) \cite{key-40}} & {\scriptsize{}Kintex 7} & {\scriptsize{}372 (slices)} & \centering{}{\scriptsize{}89} & \centering{}{\scriptsize{}6409} & \centering{}{\scriptsize{}17.228}\tabularnewline
\hline 
{\scriptsize{}ASCON-128 (LUT CRC-3) \cite{key-40}} & {\scriptsize{}Kintex 7} & {\scriptsize{}384 (slices)} & \centering{}{\scriptsize{}89} & \centering{}{\scriptsize{}6383} & \centering{}{\scriptsize{}16.224}\tabularnewline
\hline 
{\scriptsize{}ASCON-128 (unrolled) \cite{key-9}} & {\scriptsize{}Virtex4} & {\scriptsize{}26943} & \centering{}{\scriptsize{}-} & \centering{}{\scriptsize{}817.41} & \centering{}{\scriptsize{}0.031}\tabularnewline
\hline 
{\scriptsize{}ASCON-128 (recursive) \cite{key-9}} & {\scriptsize{}Virtex4} & {\scriptsize{}4021} & \centering{}{\scriptsize{}-} & \centering{}{\scriptsize{}506.29} & \centering{}{\scriptsize{}0.125}\tabularnewline
\hline 
{\scriptsize{}ASCON-128 (unrolled) \cite{key-9}} & {\scriptsize{}Virtex7 } & {\scriptsize{}22636 } & \centering{}{\scriptsize{}-} & \centering{}{\scriptsize{}1342.31 } & \centering{}{\scriptsize{}0.059}\tabularnewline
\hline 
{\scriptsize{}ASCON-128 (recursive) \cite{key-9}} & {\scriptsize{}Virtex7} & {\scriptsize{}2708 } & \centering{}{\scriptsize{}-} & \centering{}{\scriptsize{}721.53 } & \centering{}{\scriptsize{}0.266}\tabularnewline
\hline 
{\scriptsize{}ASCON-128 (unrolled) \cite{key-9}} & {\scriptsize{}Spartan6} & {\scriptsize{}22636 } & \centering{}{\scriptsize{}-} & \centering{}{\scriptsize{}688.83} & \centering{}{\scriptsize{}0.031}\tabularnewline
\hline 
{\scriptsize{}ASCON-128 (recursive) \cite{key-9}} & {\scriptsize{}Spartan6} & {\scriptsize{}2781 } & \centering{}{\scriptsize{}-} & \centering{}{\scriptsize{}346.50 } & \centering{}{\scriptsize{}0.124}\tabularnewline
\hline 
{\scriptsize{}ASCON-128a (unrolled) \cite{key-9}} & {\scriptsize{}Virtex4} & {\scriptsize{}30006 } & \centering{}{\scriptsize{}-} & \centering{}{\scriptsize{}1496.25} & \centering{}{\scriptsize{}0.049}\tabularnewline
\hline 
{\scriptsize{}ASCON-128a (recursive) \cite{key-9}} & {\scriptsize{}Virtex4} & {\scriptsize{}4215 } & \centering{}{\scriptsize{}-} & \centering{}{\scriptsize{}970.25} & \centering{}{\scriptsize{}0.231}\tabularnewline
\hline 
{\scriptsize{}ASCON-128a (unrolled) \cite{key-9}} & {\scriptsize{}Virtex7} & {\scriptsize{}25187} & \centering{}{\scriptsize{}-} & \centering{}{\scriptsize{}2419.88} & \centering{}{\scriptsize{}0.096}\tabularnewline
\hline 
{\scriptsize{}ASCON-128a (recursive) \cite{key-9}} & {\scriptsize{}Virtex7} & {\scriptsize{}2916} & \centering{}{\scriptsize{}-} & \centering{}{\scriptsize{}1357.08} & \centering{}{\scriptsize{}0.465}\tabularnewline
\hline 
{\scriptsize{}ASCON-128a (unrolled) \cite{key-9}} & {\scriptsize{}Spartan6} & {\scriptsize{}25187} & \centering{}{\scriptsize{}-} & \centering{}{\scriptsize{}1247.22} & \centering{}{\scriptsize{}0.049}\tabularnewline
\hline 
{\scriptsize{}ASCON-128a (recursive) \cite{key-9}} & {\scriptsize{}Spartan6} & {\scriptsize{}2918} & \centering{}{\scriptsize{}-} & \centering{}{\scriptsize{}638.44} & \centering{}{\scriptsize{}0.218}\tabularnewline
\hline 
{\scriptsize{}Fault-injected ASCON \cite{key-31}} & {\scriptsize{}SASEBO-GII} & {\scriptsize{}217} & \centering{}{\scriptsize{}7.8} & \centering{}{\scriptsize{}198} & \centering{}{\scriptsize{}-}\tabularnewline
\hline 
{\scriptsize{}Key-bypass HT on ASCON \cite{key-12}} & {\scriptsize{}SoC Cyclone V} & {\scriptsize{}827} & \centering{}{\scriptsize{}22.4} & \centering{}{\scriptsize{}-} & \centering{}{\scriptsize{}-}\tabularnewline
\hline 
{\scriptsize{}Round-reduction HT on ASCON \cite{key-12}} & {\scriptsize{}SoC Cyclone V} & {\scriptsize{}771} & \centering{}{\scriptsize{}22.3} & \centering{}{\scriptsize{}-} & \centering{}{\scriptsize{}-}\tabularnewline
\hline 
\end{tabular}{\scriptsize{}}}{\scriptsize\par}
\end{table*}

\begin{table*}[tbh]
\caption{Overhead results of different hardware implementations of ASCON on
ASIC hardware platform }

\centering{}{\scriptsize{}\setlength\tabcolsep{4.75pt}{}%
\begin{tabular}{|c|c|c|c|c|}
\hline 
{\scriptsize{}ASCON Architecture} & {\scriptsize{}Area ($k$GE)} & {\scriptsize{}Power ($mW$)} & {\scriptsize{}Throughput (Mbps)} & {\scriptsize{}Energy}\tabularnewline
\hline 
\hline 
{\scriptsize{}ASCON-fast (6 rounds w/ interface) \cite{key-4}} & {\scriptsize{}25.80} & {\scriptsize{}0.184 } & {\scriptsize{}13,218} & {\scriptsize{}23 $\mu J/byte$}\tabularnewline
\hline 
{\scriptsize{}ASCON-64-bit (w/ interface) \cite{key-4}} & {\scriptsize{}5.86} & {\scriptsize{}0.032 } & {\scriptsize{}72} & {\scriptsize{}1,397 $\mu J/byte$}\tabularnewline
\hline 
{\scriptsize{}ASCON-x-low-area (w/ interface) \cite{key-4}} & {\scriptsize{}3.75} & {\scriptsize{}0.015} & {\scriptsize{}14} & {\scriptsize{}5,706 $\mu J/byte$}\tabularnewline
\hline 
{\scriptsize{}ASCON-fast-TI (6 rounds w/ interface) \cite{key-4}} & {\scriptsize{}125.19} & {\scriptsize{}0.830} & {\scriptsize{}9,028} & {\scriptsize{}104 $\mu J/byte$}\tabularnewline
\hline 
{\scriptsize{}ASCON-x-low-TI (w/ interface) \cite{key-4}} & {\scriptsize{}9.19} & {\scriptsize{}0.045} & {\scriptsize{}15} & {\scriptsize{}17,234 $\mu J/byte$}\tabularnewline
\hline 
{\scriptsize{}RECO-HCON (128) \cite{key-17}} & {\scriptsize{}25.1} & {\scriptsize{}1.990} & {\scriptsize{}$5926$} & {\scriptsize{}0.335 pJ/bit}\tabularnewline
\hline 
{\scriptsize{}RECO-HCON (128a) \cite{key-17}} & {\scriptsize{}25.1} & {\scriptsize{}1.990} & {\scriptsize{}$9077$} & {\scriptsize{}0.219 pJ/bit}\tabularnewline
\hline 
{\scriptsize{}RECO-HCON (hash) \cite{key-17}} & {\scriptsize{}25.1} & {\scriptsize{}1.990} & {\scriptsize{}$3160$} & {\scriptsize{}0.637 pJ/bit}\tabularnewline
\hline 
{\scriptsize{}RECO-HCON (hash-a) \cite{key-17}} & {\scriptsize{}25.1} & {\scriptsize{}1.990} & {\scriptsize{}$4534$} & {\scriptsize{}0.439 pJ/bit}\tabularnewline
\hline 
{\scriptsize{}ASCON CMOS \cite{key-10}} & {\scriptsize{}10.1529} & {\scriptsize{}0.7459} & {\scriptsize{}-} & {\scriptsize{}478.9 pJ (64-bit)}\tabularnewline
\hline 
{\scriptsize{}ASCON CMOS \cite{key-10}} & {\scriptsize{}10.1529} & {\scriptsize{}0.7459} & {\scriptsize{}-} & {\scriptsize{}736.2 pJ (256-bit)}\tabularnewline
\hline 
{\scriptsize{}ASCON CMOS \cite{key-10}} & {\scriptsize{}10.1529} & {\scriptsize{}0.7459} & {\scriptsize{}-} & {\scriptsize{}1079.4 pJ (512-bit)}\tabularnewline
\hline 
{\scriptsize{}ASCON CMOS/STT-MRAM \cite{key-10}} & {\scriptsize{}10.7155} & {\scriptsize{}0.7148} & {\scriptsize{}-} & {\scriptsize{}427.9 pJ (64-bit)}\tabularnewline
\hline 
{\scriptsize{}ASCON CMOS/STT-MRAM \cite{key-10}} & {\scriptsize{}10.7155} & {\scriptsize{}0.7148} & {\scriptsize{}-} & {\scriptsize{}556.5 pJ (256-bit)}\tabularnewline
\hline 
{\scriptsize{}ASCON CMOS/STT-MRAM \cite{key-10}} & {\scriptsize{}10.7155} & {\scriptsize{}0.7148} & {\scriptsize{}-} & {\scriptsize{}728.1 pJ (512-bit)}\tabularnewline
\hline 
{\scriptsize{}Key-bypass HT on ASCON \cite{key-12}} & {\scriptsize{}4024 (combinational cells)} & {\scriptsize{}22.4} & {\scriptsize{}-} & {\scriptsize{}-}\tabularnewline
\hline 
{\scriptsize{}Round-reduction HT on ASCON \cite{key-12}} & {\scriptsize{}3971 (combinational cells)} & {\scriptsize{}22.3} & {\scriptsize{}-} & {\scriptsize{}-}\tabularnewline
\hline 
\end{tabular}{\scriptsize{}}}{\scriptsize\par}
\end{table*}

An FPGA-based application of ASCON cipher in portable Internet of
Medical Things (IoMT) devices is presented in \cite{key-7}, where
the cipher is used to enhance the security of such devices using AEAD
functionality. A round-based architecture of ASCON is designed for
round calculation per clock cycle. The proposed design utilizes the
dual output LUT (LUT6) feature of the Xilinx 7-series FPGA boards
to implement the $5$-bit S-box of ASCON for the optimized area. Implementing
the $5$-bit S-box of ASCON using LUT6 utilized only three LUTs in
FPGA implementation, significantly optimizing the area when compared
to other hardware implementations of ASCON. The hardware implementation
is performed on the Xilinx Artix-7 FPGA family and the area (in terms
of LUTs), throughput, frequency, and efficiency (throughput/area)
results are shown in Table 2. This proposed implementation of ASCON
cipher consumed $35\%$ less area and $56\%$ more efficiency when
compared to the architecture of ASCON in \cite{key-8}. 

In \cite{key-17}, a flexible, reconfigurable, and energy-efficient
crypto-processor to run ASCON is introduced by Wei et. al. The proposed
ASCON crypto-processor runs in six different modes: Encryption, decryption,
and hash function with different data sizes. The crypto-processor
consists of an ASCON core, shift registers (FIFO), and an I/O interface.
First, the data and text inputs are loaded into $128$-bit shift registers
as FIFO while the key, the nonce, and the target instance mode are
processed on the \textit{Start} signal. The ASCON core stays occupied
until the entire ciphertext reaches the same size as the input message
in AEAD mode, reaches $256$ bits when in hashing mode, or the tag
verification result is given out once done. The four shift registers
provide flexibility in adapting the ASCON processor to various IoT
systems with variable block sizes as they are used to divide and pad
the inputs to match the block size of the used default variant ASCON-128a.
The input sizes for other ASCON variants ASCON-128 and ASCON-hash
are processed either by splitting the $128$-bit inputs as two $64$-bit
inputs or by adapting a different counting technique to fully utilize
the space, respectively. 

The ASCON-core \cite{key-17} consists of two stages: Selective XOR
and parameterized permutation. It runs iteratively to keep the permutation
block (Section 2) busy once started. For every fixed number of rounds,
the permutation block reads a new message. Between two iterations
of the ASCON permutation, the input for the next iteration is computed
by XORing the current sponge state with either another message, key,
or a constant. The challenge
of supporting multiple instances with an arbitrary round number, variable
XOR operand, and a block size is overcome by splitting the current
sponge state into two parts: A head comprising $128$ most significant
bits (MSB), and a tail comprising $192$ least significant bits (LSB). The area, frequency, power,
throughput, and efficiency overheads of the proposed architecture
on FPGA and ASIC hardware platforms are shown in Table 2 and Table
3, respectively. 

Khan et. al. \cite{key-9} explore the hardware performance of ASCON
for artificial intelligence (AI) enabled IoT devices. Unrolled and
recursive strategies have been adopted for ASCON implementations on
Virtex-4, Virtex-7, and Spartan-6 FPGA families. The unrolled scheme
has been designed to achieve high throughput while the recursive scheme
helps in reducing hardware costs. For the unrolled architecture, the
encryption/decryption is performed using combinational circuits and
ASCON permutation is deployed in an unrolled manner for initialization,
run, and finalization phases. This results in higher throughput at
the cost of high area overhead. Moreover, since the permutation function
utilizes the same hardware for every stage, a recursive strategy is
implemented by the authors to achieve high throughput. To successfully
implement hardware re-utilization in ASCON permutation functions,
the authors propose computing two permutation rounds per clock cycle,
thus requiring a total of $24$ clock cycles for encryption/decryption
using ASCON-128 (as opposed to $36$ clock cycles for $36$ permutations)
and $26$ clock cycles for the ASCON-128a variant (as opposed to $40$
clock cycles for $40$ permutations). The area overhead is negligible
for any additional XOR operations required. The overhead results of
both unrolled and recursive implementations have been tabulated in
Table 2. 

A CMOS/STT-MRAM-based hardware implementation of ASCON benchmarked
on an ASIC hardware platform is proposed by Roussel et. al. \cite{key-10}.
Such implementation is made resilient to power failure by replacing
volatile CMOS flip-flops with non-volatile flip-flops to save the
intermediate state of ASCON computations which can then be retrieved
on startup. This hybrid CMOS/STT-MRAM implementation helps in reducing
energy utilization between $11\%$ to $48\%$, while incurring an
area overhead of about $5.5\%$ when compared to CMOS-only implementation
of ASCON. These results are presented in Table 3.

\section{Differential and Side-Channel Cryptanalysis of ASCON}

Various structural and mathematical vulnerabilities of cryptographic
ciphers with authenticated encryption such as ASCON can be exploited
to gather information from the permutation during encryption/decryption
processes. This section summarizes the works presenting cryptanalysis
of ASCON in terms of algebraic attacks, cube/cube-like attacks, differential
attacks, fault attacks, and power attacks. Most of these attacks target
the reduced round versions of ASCON to recover the secret key successfully.

\subsection{Algebraic Attacks}

In \cite{key-11}, Luo et. al. successfully attempt to recover the
entire $128$-bit key, is carried out in the software implementation,
of ASCON by attacking its permutation function and performing a soft
side-channel analytical attack (SASCA) using a factor graph method.
The factor graph of the inner permutation state is built using a template
matching technique on side-channel information leakage. Then, this
factor graph is run through their proposed Belief Propagation (BP)
algorithm to recover the secret key. From the simulations run on an
8-bit platform, it is observed that the proposed attack can recover
the entire key using a simple Hamming-weight leakage model on only
a few traces of leaked information while having low delay and memory
overheads. The authors suggest that the attack could be improved by
utilizing multi-variate value templates or machine learning algorithms.

\subsection{Cube Attacks}

Cube attacks aim to recover the secret key, one bit at a time, in
the block or stream ciphers by manipulating a set of cube variables
\cite{key-12}. The manipulated cube variables are then used to generate
encrypted messages which act as a system of linear equations which
can be solved to obtain the key bits. Any remaining key bit which
was not obtained as a solution of the linear equations can be obtained
by the brute force method \cite{key-12}. Several cube attack strategies
and implementations have been applied to the reduced-round ASCON variants.
The most successful attack has broken $7$ out of $12$ rounds of
ASCON \cite{key-19}. This section briefly compiles all the cube attacks
performed in software/hardware implementations of ASCON.

Halak and Duarte-Sanchez \cite{key-12} perform an attack on a reduced
round ASCON by mounting a cube attack on the initialization function
of the ASCON permutation in a hardware trojan (HT) compromised setting.
They leverage the vulnerabilities in the SoC FPGA hardware implementation
of ASCON-128/128a variants such as unused/ partially used states in
the FSM operations, unused values in the initial inputs, and manipulating
the round number in the iterative implementation, to inject HT in
the hardware. The authors propose two HT designs to obtain the secret
key of reduced round ASCON, namely, the key-bypass trojan, and the
round-reduction trojan \cite{key-12}. The key-bypass trojan inserts
a malicious state in the FSM which manipulates the control signals
to bypass the key directly to the output of the cipher. The round-reduction
trojan makes the cipher vulnerable to cube attack by decreasing the
number of permutation rounds from 12 to 5 in the initialization phase
of ASCON encryption, decreasing the time complexity from $2^{103.9}$
to $2^{24}$ to recover the secret key. The overhead results of HT's
compromised ASCON implementation are shown in Table 3. The area overhead
incurred is about $7\%$ when compared to the original design results
\cite{key-12}. Since the overheads of the injected trojans are low,
they go undetected in the hardware implementation. The authors propose
countermeasures such as pre-silicon circuit verification techniques
(UCI \cite{key-13}, VeriTrust \cite{key-14}, or FANCI \cite{key-15})
to detect unused inputs, post-silicon functional/structural verification
strategies, or runtime trojan detection techniques. Hardware modifications
such as pipelined/unrolled implementations, better key management,
and one-hot encoding of the FSM are also proposed to strengthen the
ASCON architecture against HTs. 

In \cite{key-19}, Li et. al. mount a practical conditional cube attack
on a reduced $5$/$6$ round ASCON. A cube-like key-subset technique
\cite{key-19} is utilized as a key dividing strategy for specific
key conditions to recover the entire key space. The practicality of
such an attack is tested in a software environment. However, the full
$12$-round ASCON implementation is resilient to this attack. The
success of this attack lies in the construction of 65-dimension cubes
due to the key dividing strategy, which divides the key into $63$
key subsets. To determine a correct subset containing a correct key
bit the \textit{cube} \textit{sum} of each cube is analyzed. This
process is continued until each of the $63$ key subsets either passes
or fails the cube tests, allowing the recovery of the secret key in
a $7$-round ASCON with a time complexity of $2^{103.9}$. The time
complexity of the proposed attack is further reduced to $2^{77}$
if the key is weak.

Similar to the work of \cite{key-19}, a practical conditional cube
attack on ASCON in a nonce-misuse setting is investigated by Baudrin
et. al. \cite{key-20}. The attack model aims to recover the capacity,
i.e., unknown inner part, of the $6$-round ASCON state right before
the encryption operation by reusing the key-nonce pair multiple times
to recover the full state and gather information about the plaintext
from corresponding ciphertexts. In \cite{key-20}, a new strategy
to search for conditional cubes in ASCON is presented, where the public
variables are split based on the coefficients of quadratic monomials
after two rounds. This allows to recover $64$ to $128$ bits of the
internal states, which in turn allows to recover the remaining $256$
bits of the state using brute force with a time complexity of up to
$2^{40}$. However, this attack does not break the security of the
original nonce respecting ASCON design \cite{key-1} and only aims
to provide insight into the potential vulnerabilities of the cipher.

The authors of \cite{key-21} also explore conditional cube attacks
on ASCON variants ASCON-128 and ASCON-128pq in a nonce-misuse scenario.
The authors implement a conditional cube attack using a proposed partial-state-recovery
method to recover $192$ bits of the $320$-bit ASCON permutation
from $2^{44.8}$ data complexity by misusing the nonce. The remaining
permutation bits are recovered using brute force in $2^{128}$ time.
After recovering all the permutation bits, the secret key is also
recovered with a time complexity of $2^{129.5}$ using $2^{31.5}$
bits of data and $2^{31.5}$ bits of memory.

A machine learning (ML)-based known plaintext attack utilizing deep
learning (DL) is proposed for the encryption operation of ASCON \cite{key-22}.
The attack predicts the plaintext with a $99.8\%$ accuracy in a nonce-misuse
setting and no finalization function. The proposed attack does not
work in a real-world application of ASCON where nonce use is respected
and a satisfactory amount of randomness is introduced in the cipher
functions so that an ML-based attack cannot match the inputs and outputs
meaningfully. 

In \cite{key-23}, Rohit et. al. perform key-recovery attacks on a
software implementation of $7$-round ASCON using a superpoly-recovery
technique called partial polynomial multiplication. Using this technique
the entire $128$-bit key is recovered from $2^{64}$ bits of data
with a time complexity of $2^{123}$, while utilizing only $2^{101}$
bits of memory. Using division properties \cite{key-24} new cube
distinguishers are also identified for $7$-round ASCON, also improving
the cube distinguishers for other reduced round ASCON implementations
to recover the secret key in a nonce/key respecting scenario.

\subsection{Differential Cryptanalysis}

In \cite{key-25}, Zong et. al. mount two differential-based collision
attack strategies on the sponge-based hash variants of ASCON namely,
ASCON-Hash and ASCON-Xof. The first attack is a non-practical kind
on a $2$-round ASCON-Hash with a time complexity of $2^{125}$. A
practical kind of collision attack is also performed on a 2-round
ASCON-Xof with a time complexity of $2^{15}$ for an output of $64$-bit.
The differential characteristics of the hash variants are found using
the MILP method and the target differential algorithm \cite{key-26}.

Gerault et. al. \cite{key-27} utilize Constraint Programming (CP)
to perform differential cryptanalysis, in software, on the permutation
of ASCON variants ASCON-128, and ASCON-128a. The capabilities of CP
in finding good differentials are used to generate differential characteristics
for ASCON to form limited-birthday distinguishers (for rounds $4$-$7$)
and rectangle attacks (for rounds $4$ and $5$). The distinguishers
are divided into black-box and non-black-box types based on their
usability for attacks on permutations with or without a key, respectively.
High-probability differentials are introduced as well to improve collision
attacks on ASCON-hash (proposed in \cite{key-25}) with a time complexity
of $2^{103}$. Additionally, multiple differential characteristics
for rounds $3$ and $4$ have been used forgery attacks, in a nonce-respecting
setting, on the permutation and finalization functions of reduced-round
ASCON-128 and ASCON-128a. The time complexity for such an attack for
the three rounds of finalization operation is determined to be $2^{32.76}$.
For 4-round ASCON-128, the time complexity of such an attack in the
finalization phase is $2^{96.61}$. However, in this scenario, the
attack exceeds the recommended amount of data blocks processed for
a single key. For ASCON-128a, the forgery attack is performed on both
the iterative permutation and the finalization functions. For the
iterative permutation, the time complexity for the proposed attack
is $2^{117}$, however, it also exceeds the processed data blocks
limit for a given key established by the ASCON designers \cite{key-1}.

Using the undisturbed bits in the ASCON S-box, Tezcan \cite{key-28}
performs a truncated, impossible, and improbable differential analysis
on reduced $4$/$5$ round ASCON. It is observed that there are $35$
undisturbed bits in the execution of the S-box which are used to generate
the aforementioned differentials. The impossible differential analysis
is based on the idea that for a specific difference, the differential
cannot occur, i.e., its probability is zero. This helps in finding
the correct key by removing the incorrect keys determined using impossible
differences. The key recovery attack is performed on $4$/$5$ rounds
of ASCON using truncated differentials which have a probability of
$1$. Truncated differentials with a probability of 1 are coupled
with the cipher symmetry and can help determine whether the two key-bits
associated with the active S-boxes are $1$. Thus, for a truncated
attack mounted on a $4$-round ASCON, 16 key bits would be 1 and the
remaining $48$ key bits can recovered by brute force using $3^{48}$
encryptions in $4$-round ASCON using $2^{2}$ bits of data. A truncated
attack on $5$-round ASCON can recover $70$ key bits and the remaining
key bits are recovered in either $2^{58}$ encryptions (for weak keys)
or $2^{128}-2^{64}$ encryptions with $2^{109}$ bits of data. The
improbable differential attack on $5$-round ASCON is similar to the
aforementioned truncated differential attack which again uses $2^{109}$
data bits to determine the ASCON permutation from a random permutation
by complementing the output differences. The miss-in-the-middle technique
is combined with the truncated differential in the decryption operation
to mount an impossible differential attack on $5$-round ASCON by
using only $2^{256}$ data bits.

In \cite{key-29}, Hu et. al. perform probabilistic and deterministic
high-order differential/differential-linear (HD/HDL) attacks on a
software implementation of ASCON. The probabilistic HD/HDL attack
is performed using a higher-order algebraic transitional form function
(HATF) technique which is helpful for the cryptanalysis of quadratic
round functions. Using HATF, various highly-biased $2nd$-order HDL
approximations are discovered for the initialization function of reduced
$5$-round ASCON up to eighth order. Thus, using HATF, the key-recovery
attack on $5$-round ASCON can be performed with a complexity of $2^{22}$,
and the distinguishing attack on reduced-round ASCON can be performed
with a complexity of $2^{12}$. A conditional $3rd$-order HDL approximation
is also proposed for the initialization function of $6$-round ASCON.
High bias bits are observed in the S-box in the 6th round of ASCON
using the HATF with $24$ conditions with a theoretical bias value
of $2^{-22}$ \cite{key-29}. This value is computed using the bias
observed in the bits of round-$5$ S-box ($2^{-14}$), pilling-up
lemma, and $8$ HDL approximations applied to $2^{30}$ test samples.
The deterministic HD attack is also performed using the differential
support function (DSF) technique, that helps in finding HD distinguishers
by performing efficient linearizations on permutation inputs. Thus,
the DSF technique improves the complexity of a distinguishing attack
on the permutation of reduced 8-round ASCON from $2^{130}$ to $2^{46}$.
Using a similar DSF method, the zero-sum distinguishers for $12$-round
ASCON permutation are calculated with a time and data complexity of
$2^{55}$, respectively. \cite{key-29}.

\subsection{Fault Analysis}

Ramezanpour et. al. \cite{key-30} successfully apply a statistical
ineffective fault analysis (SIFA) using double-fault injection and
key-dividing techniques on the S-box of ASCON in a software implementation
to recover the secret key. The authors inject faults in any selected
pair of S-boxes for every encryption performed in the last round of
the finalization stage of ASCON. The faults are injected using a clock
and/or voltage glitches are injected in a manner where they do not
affect the result of the S-box. Then, the correct tag values resulting
from induced ineffective faults are analyzed to gather information
about the secret key. The probability of distribution is assumed to
be biased in the proposed attack and the attack is successful as long
as there is sufficient data available for analysis ($12.5$ to $2500$
correct tag values in the proposed study). Thus, the SIFA-based fault
model requirements are less than other differential fault analysis
techniques and are also noise tolerant. Fault attack countermeasures
such as error detection and error-randomization techniques fail in
the presence of SIFA as they rely on the incorrect output value in
cipher operations under fault inductions. The best countermeasures
against SIFA are those where the fault injection mechanisms can be
detected or where the fault distribution is independent of secret
data. Sensor-based techniques can detect fault injection mechanisms,
however, they are limited in the fault mechanisms they can detect.
The FPGA and ASIC hardware implementations are proposed as future
work.

Surya et. al. \cite{key-31} implement a synchronous clock glitching
strategy to induce delay faults in selected parts of the ASCON architecture.
SASEBO-GII FPGA board is used for the hardware implementation and
a Digital Clock Manager (DCM) is used to generate synchronous clock
signals. In every encryption round, the faults are injected into the
ASCON S-box via a high-frequency faulty clock signal resulting in
faulty output in the ASCON linear layer. Surya et. al. also try to
implement an asynchronous clock glitching strategy, however, it incurs
higher utilization overhead when compared to the synchronous method
\cite{key-31}. The fault injection is performed by feeding the faulty
clock signal only to a few parts of the design to better observe the
error distribution and propagation, and to emulate EM injections easily.
Possible countermeasures to this kind of attack include a threshold
implementation (TI) scheme and a unified masking approach to protecting
the ASCON architecture. Hardware implementation overheads are given
in Table 2. 

Joshi and Mazumdar \cite{key-32} perform a subset fault analysis
(SSFA) on a software implementation of ASCON-128 by attacking the
vulnerabilities in its S-box. The strategy tries to find correlations
between the input and output bits of the ASCON S-box by determining
which output bits become 0 for input bits set to $0$. They also propose
a key division strategy to decrease the search space for key recovery
to $2^{64}$ for the worst case. Key masking before key whitening
operation, error detection via partial decryption, or using a new
and strengthened S-box design resilient to 1-bit SSFA are proposed
as the countermeasures against SSFA. 

In \cite{key-32}, a fault attack called the preliminary attack which
focuses on vulnerabilities in the key whitening function and the tag
creation function of the finalization stage is also proposed. Both
attacks are shown to retrieve the entire secret key using a key analysis
methodology. For the preliminary attack strategy, the involution property
of the XOR function is leveraged to recover the key value from the
generated tag. The attack can be mounted on ASCON in three ways: 1)
Injecting faults into three selected S-boxes which requires $374$
fault injections to recover the full secret key; 2) Injecting faults
into a single selected S-box followed by an instruction skip error
(induced by another fault injection) which requires $256$ fault injections
to recover the entire secret key; and 3) Resetting a word register
to 0 at the output of the substitution layer via $128$ (for $1$-bit
faults), $16$ (for $1$-byte fault), or $2$ (for $64$-bit word
fault) fault injection, respectively, to recover the entire key. 

Ramezanpour et. al. \cite{key-33} propose another statistical fault
analysis attack called the fault intensity map analysis (FIMA) for
a software implementation of ASCON. This attack can retrieve the entire
$128$-bit key of ASCON. The attack is designed to use different features
such as faulty ciphertexts, SIFA-induced correct ciphertexts, and
data-dependent bias to recover the secret key. It is also resilient
several countermeasures such as error detection techniques for DFA
where it can gather secret information from the increased sample size.
Even with infective countermeasures, where a fault is injected in
a wrong random round of ASCON, FIMA can recover the secret key with
$453$ data samples. Thus, compared to other fault analysis methods,
FIMA is $6$ times more powerful. 

Ambili and Jose \cite{key-34} propose an upgraded design of ASCON-128a
using pseudo-randomness of Cellular Automata (CA) to make the cipher
resilient against SIFA and SSFA, verified mathematically in a software
implementation. CA is a technique where a particular cell updates
its value every iteration depending upon its state and a set of predefined
rules. In \cite{key-34}, a null boundary CA (where the farthest cell
neighbor is set to $0$) is used to protect the architecture of ASCON
by implementing it in the pseudorandom function of ASCON permutation.
The security against SIFA and SSFA is due to the induced randomness
in the linear layer due to which the XOR/linear equations derived
for erroneous bits cannot be solved reliably. The authors propose
the practicality of their work in hardware implementation as future
work. 

Kaur et. al. \cite{key-40} propose low-cost error-detection mechanisms
as countermeasures against fault attacks for the hardware implementations
of ASCON. Parity, interleaved parity, and CRC-3-based techniques are
formulated and applied to the $5$-bit S-box of ASCON to detect natural
and transient faults injected that may occur during the S-box operation
to generate faulty outputs. Two kinds of error-detection implementations
are introduced, either using Boolean logic or using the Look-up Tables
(LUTs). The error coverage of the proposed error detection schemes
is tested for $640,000$ injected faults and is determined to be over
$99.99\%$ \cite{key-40}. The overhead results for the hardware implementations
of ASCON on Spartan-7 and Kintex-7 FPGAs, protected using these error-detection
mechanisms, show an increase in the area overhead up to $15\%$ for
both types of implementations. The results have been tabulated in
Table 2. These mechanisms aim to detect most of injected single and
multiple-bit faults leveraged in DFA and SSFA, however, detecting
SIFA could be challenging since the error detection is performed at
the output of the S-box and not the input. 

\subsection{Power Analysis}

The research work of \cite{key-35} introduces a machine learning
(ML) based side-channel analysis with reinforcement learning (SCARL)
to obtain confidential data by using unsupervised learning to extract
leakage models from power measurements. SCARL attempts to obtain the
secret key by analyzing the power consumed by the non-linear S-box
computations in the initialization phase of ASCON. An autoencoder
to process power measurement samples and reinforcement learning along
with actor-critic networks are used to cluster the power features.
The hardware implementation of the ASCON-128 on the Artix-7 FPGA board
is attacked using SCARL, where FOBOS is used to gather the power measurements
of $64$ S-box computations. The authors successfully demonstrate
that their proposed SCARL strategy can recover the secret key of the
implemented ASCON-128 cipher on Artix-7 FPGA by using power measurements
obtained during $24,000$ encryption operations; the first $4$ bits
of the secret key are obtained using SCARL within $8$ minutes.

The SCA countermeasure assessment based on power leakage is performed
using FOBOS 2 and is presented in \cite{key-6}. Test vector leakage
assessment (TVLA) results for protected and unprotected ASCON architecture
using Artix-7 and Spartan-6 FPGA boards. Significant leakage is noticed
in the unprotected version vs. the protected version in which the
values are within the threshold value; however, no confidential data
is recovered through power leakage. The $\chi^{2}$-test is also performed
for the protected and unprotected ASCON architecture for leakage assessment
flow for fixed and random frequency classes using the same test vectors
as that for TVLA. Similar to TVLA, the $\chi^{2}$-test observes leakage
in the unprotected ASCON while no leakage is observed in the protected
version.

ASCON's initialization phase is the most vulnerable to power analysis
attacks as only $2$ input bits (out of $5$) of the initial S-boxes
are unknown (secret key bits) while the other $3$ are known. In \cite{key-4},
in addition to optimized hardware implementations of ASCON, the authors
propose countermeasures against side-channel analysis attacks, particularly
first-order differential power analysis (DPA) attacks. This is achieved
by using the TI scheme \cite{key-36}-\cite{key-37}, a masking technique
where the calculations on critical data are indirectly carried out
by modified transformations called shares. The proposed protected
implementation of ASCON efficiently applies three shares like in the
Keccak since ASCON uses an affine transformation of Keccak's $\chi$
function \cite{key-37}. ASCON's linear layer can be implemented on
each share. However, the non-linear S-box layer needs to be transformed
such that it maintains the following properties 1) Correctness - the
sum of the resulting output share matches the S-box output when applied
to the sum of the input shares, 2) Non-completeness - each of the
three S-box functions is independent of at least one input share,
and 3) Uniformity - each S-box function is invertible. The research
work \cite{key-4} also implements a three-share TI version of the
ASCON-fast and ASCON-x-low-area variants. The ASCON-fast-TI is a microcontroller-based
implementation proposed as a cryptographic co-processor where the
initial state sharing and randomness are performed by the microcontroller.
The ASCON-x-low-TI variant directly uses the output of an available
random number generator in the S-box operation per cycle. The hardware
implementation results of the TI-protected implementation are listed
in Table 2. 

In \cite{key-38}, the correlation power analysis (CPA) and DPA attacks
are mounted on the parallel implementations of ASCON-128 and the TI-protected
ASCON-128, respectively. The CPA attack is successfully implemented
on ASCON-128 by attacking the vulnerabilities at the end of the initialization
phase while requiring fewer power traces to obtain half of the secret
key. These vulnerabilities in the initialization round are leveraged
again for the attack on the TI-protected ASCON-128 by using the difference
of skewness as the third-order attack \cite{key-38}.

Similarly, the research work of \cite{key-39} implements a TI-based
protection scheme for ASCON against first-order DPA by executing one
round in $7$ clock cycles. The implemented scheme instantiates a
single hybrid $2$-share/$3$-share TI-protected $64$-bit AND module
which uses random $192$ bits every clock cycle - $128$ for resharing
between $2$-/$3$- shares, and the remaining $64$ bits are used
to maintain TI uniformity. This TI-protected ASCON implementation
is shown to be resistant to first-order DPA by analyzing the results
of the $t$-test leakage detection test \cite{key-39}.

\section{Our Insights, Visions, and Conclusion}

This survey is the first work on the current standard for lightweight
cryptography, standardized in 2023. This study covers various hardware
implementations proposed for NIST LWC winner ASCON in the recent years
on FPGA and ASIC hardware platforms. These hardware implementations
suggest improvements on the original design in terms of area, throughput,
efficiency or energy/power utilizations for applications of ASCON
in resource-constrained devices. Differential and side-channel cryptanalysis
performed on ASCON on both hardware and software platforms have also
been reviewed. The differential cryptanalysis techniques highlight
the vulnerabilities present in the permutation function of the reduced-round
ASCON but not of a full $12$-round ASCON. The S-box design is also
shown to be vulnerable to SFA due to its design where the secret key
is retrieved using correlation between the input and the output bits.
Power analysis attacks were also mounted on ASCON using machine learning
strategies or DPA to gather secret information from side-channel leakage,
however, protected architecture of ASCON using TI is considered safe
against such attacks. 

An insight here is to investigate augmenting the ASCON implementations
with design-for low-cost fault diagnosis and have that as a design
decision factor. The merit of taking low overhead into account before
designing cryptographic algorithms is that the resulting ASCON architectures
will be designed for low-overhead error detection and the countermeasures
are not devised aftermath.

One other interesting insight/vision related to fault and side-channel
attacks is combined attacks and countermeasures. We believe this would
be the future of such attacks for ASCON (there has been little prior
work and none considers ASCON \cite{key-61}-\cite{key-64}).

\section*{Acknowledgments}

This work was performed under the U.S. federal agency award 60NANB20D013
granted from U.S. Department of Commerce, National Institute of Standards
and Technology (NIST).

\end{document}